\documentstyle[prb,preprint,aps]{revtex}
\begin{document}
\draft
\title{Anisotropy of zero-bias diffusive anomalies for different orientations
of an external magnetic field}

 \author{E.V. Sukhorukov, A.V. Khaetskii}
 \address{Institute of Microelectronics Technology, Russian Academy of Sciences,
 142432, Chernogolovka, Moscow district, Russia}
\maketitle

\begin{abstract}
We consider the influence of the electron-electron interaction on the
nonlinearity of the current-voltage characteristic of the tunnel junction at low
bias (diffusive anomaly) in the presence of the classical magnetic field.
We present the theory of a new phenomenon which manifests itself in
the strong anisotropy of a diffusive anomaly for different orientations of
the magnetic field with respect to the interface of the tunnel junction.
The nonlinear differential tunneling conductance has a universal
magnetic field dependence, so that only the magnetic field component
perpendicular to the interface
is involved.
In particular, when the magnetic field is
parallel to the interface, the I-V characteristic does not depend on the value of
the magnetic field.
\end{abstract}

\section{Introduction.}
\par
As it was shown by Altshuler
and Aronov
  \cite{Alt1},
the electron-electron interaction
strongly influences various physical properties of electron systems. Among those
are conductivity, spin susceptibility and the one-particle density of states at
the Fermi level.
Suppression of the one-particle density of states at the Fermi level
leads to appearance of small deviations from Ohm's law in the
current-voltage characteristics of a tunnel junction at small voltages
$V$
(zero-bias diffusive anomalies). These authors used the diagrammatic perturbative
method.  A more complete description
of the
nonlinearities
in the I-V characteristics of
tunnel junctions
which allows to go beyond the perturbative
treatment
was given in Refs.(\onlinecite{Naz1,Naz2,Naz3,Levit}). (Earlier the
non-perturbative treatment of renormalization of one-particle density of states
due to electron-electron correlations was given in Ref. \onlinecite{Finkel}).
It was shown that the
strength
of the diffusive anomaly can be characterized by the effective
frequency
dependent impedance ($\omega\sim eV/\hbar$)  which
the electron feels when tunneling. Immediately after the electron tunnels,
the system acquires an extra energy due to the interaction between this
electron and the electrons of the liquid. Therefore, the electron density
perturbation must spread in order to reach the final state. In this process
the many-body system acquires some action which is proportional to the impedance
mentioned above.
The value of this
effective impedance can be estimated as the resistance of the metal piece which
the electron covers in
time $\sim\hbar /eV$ during its diffusion motion after the tunneling.
The value of this impedance
governs the magnitude of the deviations from Ohm's
law. When the conductivity of
the system is quite large,  the effective
impedance is much smaller than the quantum unit of resistance $h/e^2$
and, as a result, there is only a small deviation from the Ohmic behavior.
In the present work we will assume that this condition is satisfied.
The classical magnetic field influences the charge spreading (through the
conductivity), hence, the zero-bias anomaly depends on the magnetic field
value.
\par We present here the theory of a new phenomenon which manifests itself in
the strong anisotropy of a diffusive anomaly for different orientations of the classical
magnetic field with respect to the current direction. For definiteness,
we considered the tunnel
contact of two 3D semi-infinite samples and the magnetic field vector can have
an arbitrary angle with the  interface plane (see Fig.\ \ref{1}).

As is common, we consider  the case of a metallic regime
($\epsilon_{\scriptscriptstyle F}\tau_p\gg 1$, $\tau_p$ is the momentum
relaxation time).
We have solved the equations of classical
electrodynamics which describe the propagation of the charges and electric
fields for this geometry and obtained the following results.
The nonlinear differential tunneling conductance has a universal
magnetic field dependence, so that only the magnetic field component
perpendicular to the interface is involved.
This is related to the fact that the charge
spreading after the tunneling has a completely different character for the classical
magnetic field  oriented perpendicular or parallel to the interface.
Namely, the magnetic field perpendicular to the interface strongly reduces the
transverse conductivity $\sigma_{xx}$ of the system (in the region of classically
strong magnetic fields), hence, the spreading of the charge along the
interface plane is strongly blocked. The effective impedance increases and has
a universal dependence on the magnetic field $[1+(\omega_c\tau_p )^2]$
(multiplied by the impedance value
for a zero magnetic field). This means that the
nonlinear tunneling conductance
also has this universal magnetic field dependence. For
the magnetic field parallel to the interface the charge spreading along the
interface
plane is not blocked because this spreading easily occurs through the Hall
drift along the interface.
As a result, the I-V characteristic for this
orientation of the magnetic field does not depend on the magnetic field value.
This phenomenon resembles the absence of magnetoresistance in the metal at low
temperatures for one group of carriers. In the case of arbitrary orientation
only the perpendicular component of the magnetic field is involved because the
contribution of the parallel component is canceled by the Hall electric fields.

\par The size effects in one-particle density of states in the presence of 
electron-electron
correlations were studied in Ref. \onlinecite{Alt3}. This consideration was restricted
to the case of a zero magnetic field. In contrast, the influence of the classical
magnetic field on the electron density of
states was studied only for the case of bulk metal
\cite{Alt2}. The authors of the latter did not take into account the
important role of the surface of a metal in this phenomenon and could not
obtain the strong dependence of the zero-bias anomaly on the orientation
of an external magnetic field.

\par The results of present work
explain the observations of the recent experiment
\cite{Dub}. In the latter the magnitude of the diffusive anomaly for
the electron tunneling between two heavily doped GaAs samples
was measured for two different orientations of the external magnetic field parallel
and perpendicular to the interface. It was observed that the magnitude of the diffusive
anomaly does not depend on the magnetic field value in the case when the magnetic
field is parallel to the interface and strongly depends on it in the case of
perpendicular orientation.

\section{Formalism.}
\par
Let us consider the concrete geometry of the experiment.
We assume that the leads are two 3D semi-infinite conducting
samples isolated by a thin high barrier of  width $d$
 (see Fig.\ \ref{1}).
If a negative voltage is applied to the left lead, the electron may tunnel
through the barrier from the left lead to the right one leaving a
hole behind.
The magnetic field applied governs the charge spreading process after
tunneling. In general, the magnetic field is arbitrary
oriented so that the diffusion coefficients $D_{km}$ depend not only on
the magnetic field value, but also on the  angle $\theta$ between the
magnetic field and the plane of the insulating layer.

\par We use the tunnel Hamiltonian method. The total Hamiltonian of the problem
consists of the following parts:
\begin{equation}
\hat {H}=\hat {H}_R+\hat {H}_L+\hat {H}_{imp}+\hat {H}_T\;.
\label{1}
\end{equation}
Here $\hat {H}_{R,L}$ describes the electron system in the right and the left
leads with the electron-electron interaction taken into account. The Hamiltonian
$\hat {H}_{imp}$ describes the electron scattering by the impurities and the tunnel
term $\hat {H}_T=\hat {A}+\hat {A}^{\dagger}$ describes the transitions of
 electrons through the insulating layer,
\begin{equation}
\hat {A}=\int\limits_{\stackrel{x > d/2}{\scriptscriptstyle x^{\prime}< -d/2}}
d^3{\bf r}d^3{\bf r}^{\prime}T({\bf r},{\bf r}^{\prime}
)\psi^{\dag}({\bf r})\psi ({\bf r}^{\prime})\;.
\label{2}
\end{equation}
The corresponding expression for the current operator is
\begin{equation}
\hat I=i\left(\hat A-\hat A^{\dag}\right)\,.
\label{3}
\end{equation}
\par As usual, the current is calculated in the second-order perturbation theory in
the tunnel Hamiltonian. For  arbitrary voltage $V$ the current can be
obtained \cite{Mahan} by the
analytical continuation of the following Matsubara Green's function determined
at $\omega_n=2\pi Tn$ ,
\begin{equation}
I(V)=2e Im\left\{\int\limits_{0}^{\beta}d\tau\:e^{i\omega_n\tau}\left
.\left<\hat T\hat A(\tau )\hat A^{\dag}(0)\right>\right|_{i\omega_n\to
-eV+i0}\right\}\;.
\label{4}
\end{equation}
From now on we assume that $e >0$ and $\hbar =1$.
This Green's function is calculated in the standard fashion by
averaging over the electrodynamic field configurations \cite{Naz2}
and can be expressed in terms of the effective action
which describes the low-frequency electrodynamic modes of the system
excited during the tunneling.
When the action is small ($S(\tau )\ll 1$),
one can get the following simple expression for the small correction
to the current-voltage characteristic at $T=0$:
\begin{equation}
R_t{{\partial^2I}\over {\partial V^2}}=
{{2e}\over {\pi}} Jm\left\{\left.S(\omega
)\right|_{i\omega\to -eV+i0}\right\}\;,
\label{5}
\end{equation}
where $R_t$ is the tunnel resistance of the junction.
\par At low frequencies
($\omega\tau_p\ll 1$,
$\tau_p$ is the momentum relaxation time),
when the diffusion approximation is
valid, one can get the following formula for the action \cite{Naz2}:
\begin{equation}
S(\omega )={1\over 2}\int\limits d^3{\bf r}\;\rho_{-\omega}({\bf r})
\phi_{\omega}(
{\bf r}),\;\;\;
\label{6}
\end{equation}
where $\phi_{\omega}({\bf r})$ is the electrodynamical
potential excited by the tunneling electron, and
$\rho_{\omega}({\bf r})=\int\limits_{-\infty}^{+\infty}
d\tau\exp (i\omega\tau )\rho ({\bf r},\tau )$
is the Fourier transformation of the probability density to find the electron
(hole) at point ${\bf r}$ if ${\bf r}_0$ is the point of tunneling on the
interface of the tunnel junction. This quantity is just a common electron-hole
(diffusion) propagator which stands in usual diagrammatic approach \cite{Alt1}.
$\rho_{\omega}({\bf r})$ obeys the
diffusion equation in imaginary time  in the presence of the magnetic field
(different signs correspond to different leads),
\begin{equation}
|\omega |\rho_{\omega}({\bf r})-\sum_{k,m=x,y,z}D_{km}\nabla_k\nabla_m
\,\rho_{\omega}
({\bf r})=\mp e\:sign(\omega )\delta ({\bf r}-{\bf r}_0)\:,
\label{7}
\end{equation}
with the boundary conditions that there is no diffusive current across the
surfaces of the leads:
\begin{equation}
\sum_kD_{xk}\nabla_k\,\rho_{\omega}({\bf r})\left|_{x=\pm d/2}=0\right
.
\label{8}
\end{equation}
This boundary condition is imposed at the distance from the surface
which is 
larger than mean free path. It should be noted that in the case of
$\omega_c\tau_p\gg 1$ the diffusion coefficients are renormalized due to
the existance of skipping orbits at  the distance of the order of cyclotron radius
$R_c$ from the surface. However, in the diffusion approximation we consider here  the contribution of
the sliding current to the action $S(\omega)$ is small with respect to the
parameter $qR_c \ll 1$, where $q^{-1}$ is the characteristic length scale in the
problem.\cite{Maslov}
\par
In the coordinate system when the magnetic field is parallel to the
$z$-axis, the
diffusion coefficients have the following form
($\omega_c$ is the cyclotron frequency):
\begin{equation}
\tilde D_{zz}=D_0={{v_{\scriptscriptstyle F}^2\tau_p}\over 3}\;,\;
\;\;\tilde D_{xx}=\tilde D_{yy}
={{D_0}\over {1+\omega_c^2\tau_p^2}}\;,
\label{9}
\end{equation}
\begin{equation}
\tilde D_{xy}=-\tilde D_{yx}= -sign(\omega )\,D_0\,{{\omega_c\tau_p}\over{1+\omega_
c^2\tau_p^2}}\;.
\label{10}
\end{equation}
For an arbitrary orientation of the magnetic field in the x-z plane the
diffusion coefficients
 can be obtained from the diffusion
matrix $\tilde D$ (Eqs.\ (\ref{9},\ref{10})) by the rotation over the angle $\theta$
around the $y$-axis: $D(\theta)=\hat U^T(\theta)\tilde D \hat U(\theta)$.

\par We assume that the conductivity $\sigma $ of the leads is large, so for the frequencies
in question ($\omega\ll \sigma$) we use the electroneutrality  condition
(Debye radius $r_{\scriptscriptstyle D}=0$) when calculating the electrodynamical
potential
$\phi_{\omega}({\bf r})$. In this case there
are no charges in the 3D leads except the 2D charges
$
\pm Q_{\omega}({\bf r}_{\parallel})
$
on their surfaces,
and $\phi_{\omega}({\bf r})$ obeys the
system of the following equations in imaginary time:
\begin{equation}
\sum_k\nabla_kj_k(\omega ,{\bf r})=0\,,
\label{11}
\end{equation}
\begin{equation}
j_k(\omega ,{\bf r})=-\sum_m\sigma_{km}\nabla_m\phi_{\omega}({\bf r})\,,
\label{12}
\end{equation}
\begin{equation}
\left.j_x(\omega ,{\bf r}_{\parallel})\right|_{x=\pm d/2}=-|\omega |Q_{\omega}
({\bf r}_{\parallel})-e\,
sign(\omega )\,\delta ({\bf r}_{\parallel})\,.
\label{13}
\end{equation}
Conductivity is related to the diffusion tensor
through the Einstein relation, $\sigma_{km}=e^2\nu_{\scriptscriptstyle F}\,D_{km}\,,$
where $\nu_{\scriptscriptstyle F}$ is the
density of states for noninteracting electrons at the Fermi surface.
Outside the leads $\phi_{\omega}({\bf r})$ obeys the Laplace equation.
Since for the small frequencies considered here the characteristic
spatial scale of the variation of $\phi_{\omega} $ inside the leads
is much larger than $d$, we can neglect the electric fields in the leads
while finding the relation between  values of $\phi_{\omega}$ and
$Q_{\omega}$ at the surfaces. Hence, we obtain a simple relation
\begin{equation}
\left.\phi_{\omega}({\bf r}_{\parallel})\right|_{x=\pm d/2}
= \pm
2\pi d Q_{\omega}({\bf r}_{\parallel})/\epsilon_i \;,
\label{14}
\end{equation}
where $\epsilon_i$ is the dielectric constant of the insulating layer.
\par  It should be noted that according to Refs.(\onlinecite{Naz1,Naz2,Naz3}) 
the value $Z(\omega )=e^{-2}\omega S(\omega )$
can be thought as the
effective frequency dependent impedance which the electron feels
when tunneling.
\par  Finally, we would like to mention the connection between
the method  used in the present paper and the diagrammatic approach \cite{Alt1}.
Calculation of the correlator in Eq.(\ref{4}) corresponds (in the lowest order) to the account of diagrams
a) and b) in Fig.\ \ref{2}.\cite{Naz3}
Diagram a) is the exchange correction to one-particle density of states in
each lead. Diagram b) contributes to the correlator due to excitation of the 
gapless interface plasmon during electron tunneling. We do not take into account
Hartree diagram c) which is small \cite{Alt1} in parameter
$\lambda_{\scriptscriptstyle F} /r_D \ll 1$ because we consider the case
of the weakly non-ideal electron gas.

\section{Calculation of action $S(\omega )$.}
\par
 Differential Eq.\ (\ref{7}) with  boundary
condition (\ref{8}) can be easily solved by the Fourier transformation with
respect to the $y$ and $z$ coordinates along the surface plane.
Then, for the charge density in the right lead
($x>d/2$) we obtain
\begin{equation}
\rho_{\omega}(x,{\bf q})=-{{e\,sign(\omega )\,}\over {\Delta D_{xx}
-iq_yD_{xy}}}\:
\exp\left[-\left(iq_zD_{xx}^{-1}D_{xz}+\Delta\right)\left(x-d/
2 \right)\right],
\label{15}
\end{equation}
where $\Delta^2=D_{xx}^{-1}\left[|\omega| +D_{yy}q_{y^{}}^2+
\left(D_{zz}-D_{xx}^{-1}D_{xz}^2\right)
q_z^2\right]$.
A similar procedure with  Eqs.\ (\ref{11}-\ref{14}) for the electrodynamic
potential $\phi_{\omega}$ gives
\begin{equation}
\phi_{\omega}(x,{\bf q})=-{{e\,sign(\omega )}\over {(2\pi d)^{-1}\epsilon_
i|\omega |+\delta\sigma_{xx}-iq_y\sigma_{xy}}}\,
\exp\left[-\left(iq_z\sigma_{xx}^{-1}\sigma_{xz}+\delta\right)\left(x-d/
2\right)\right],
\label{16}
\end{equation}
with
$\delta^2=\sigma_{xx}^{-1}\left[\sigma_{yy}q_{y^{}}^2+
\left(\sigma_{zz}-\sigma_{xx}^{-1}\sigma_{xz}^2\right)
q_z^2\right]$.
Then, for the
characteristic frequency scale of the problem ($\omega \sim Dq^2$,
$q\ll l_i^{-1}$ and $l_i=v_{\scriptscriptstyle F}\tau _p$) the
term $(2\pi d)^{-1}\epsilon_i|\omega |$ is small compared to others in
the denominator
of the right hand side of Eq.\ (\ref{16}), provided
$d\gg r_{\scriptscriptstyle D}^2/l_i$
($r_{\scriptscriptstyle D}^2=D_0/\sigma_0$).
The last inequality is usually satisfied for tunnel junctions, and
we omit this term  in the expression for $\phi_{\omega}$
in further calculations. This approximation actually means that the
spreading of the charges in the left and right leads occurs absolutely
independently. It is determined by the diffusive modes
and is not effected by surface plasmon modes which are not excited
in the frequency range considered (it means that we neglect the contribution of
diagram b) in Fig.\ \ref{2}).

\par By inversion of the coordinates and the sign of the electron charge in
Eqs.\ (\ref{15}, \ref{16}) we can obtain the potential and the charge density
in the left lead. Then, $S(\omega )$ in Eq.\ (\ref{6}) is given by
the integral over the right lead multiplied by $2$.
Carrying out the integration over $x$ and over ${\bf q}$,
we present $S(\omega )$ in the following universal form:
\begin{equation}
S(\omega )=S_0(\omega )F(H,\theta ),\quad S_0(\omega )=-{{e^2}\over {
2\pi\sigma_0\sqrt {|\omega |D_0}}}\;, \quad
F(H,\theta )=1+\omega_c^2\tau_p^2\sin^2\theta\:.
\label{17}
\end{equation}

\section{Nonlinear differential conductance.}
\par
In the expression for $S(\omega$) (see Eq.\ (\ref{17})) the dependence on
$\omega$ is
extracted in  factor $S_0(\omega )$, i.e. the action for $H=0$.  Factor
$F(H,\theta )$ depends only on the magnetic field. Hence, to obtain the
correction to the current-voltage characteristic we can easily perform
the analytical continuation in  Eq.\ (\ref{17}),
\begin{equation}
R_t{{\partial^2I}\over {\partial V^2}}={{e^3}\over {\pi^
2\sigma_0\sqrt {2eVD_0}}}\:F(H,\theta )\,.
\label{18}
\end{equation}
It seems to be more convenient to represent this result in terms of
the correction $\delta R(H,\theta )$ to the Ohmic resistance of the tunnel junction.
To do this, we integrate the above equation over $V$,
\begin{equation}
R_t{{\partial I}\over {\partial V}}=\left[{{e^2}\over {\pi^2\sigma_
0}}\left({{2eV}\over {D_0}}\right)^{1/2}-{{\delta R(0)}\over {R_t}}\right
]F(H,\theta )\:,
\label{19}
\end{equation}
so that the correction to the resistance at zero bias is
$\delta R(H,\theta )=\delta R(0)F(H,\theta )$, and $\delta R(0)$ can be estimated
by comparison of the two terms in Eq.\ (\ref{19}) at $eV\sim 1/\tau_p\;$:
$\delta R(0)/R_t\sim (\lambda_{\scriptscriptstyle F}/l_i)^2$.

 Function $F(H,\theta )$ governs the dependence of the zero-bias anomaly
on the magnetic field. As it follows from
Eq.\ (\ref{17}),
there is a strong anisotropy of the diffusive anomaly with respect to the
orientation of the magnetic field. In particular, if the magnetic field is
exactly parallel to the interface,
then
$F(H, 0)=1$ and the correction to the resistance of the tunnel junction
does not depend on the magnetic field value.
In our opinion, this quite interesting effect has the same physical origin
as the absence of the magnetoresistance in the metal for one group of
carriers. Namely, additional spreading of the charge along the interface
plane  due to the Hall electric fields that appear in this geometry
cancels exactly the blocking of the charge spreading due to decreasing  $\sigma_{xx}$
in the magnetic field. In contrast, for the case when the magnetic field
is perpendicular to the interface plane, the Hall electric fields do not
occur and the charge spreading is governed only by $\sigma_{xx}$. As a result,
for $\theta=\pi /2$
we have
$F(H,\pi /2)=1+\omega_c^2\tau_p^2$, which is just the magnetic field
dependence of $\sigma_{xx}^{-1}$.
It is worth  mentioning that in general  factor $F(H,\theta)$ depends
only on the component of the magnetic field $H_x$ 
perpendicular to the interface.

\par In conclusion, we have considered the influence of the classical magnetic field
on the diffusive anomaly in the I-V characteristic of the tunnel junction.
We have shown that the nonlinear differential tunneling conductance strongly depends on 
the orientation of the magnetic field
with respect to the interface plane. This phenomenon is explained by the
different character of charge spreading after  tunneling for
different orientations of the magnetic field. The result of this
work explains the observation of the recent experiment \cite{Dub}.

This work was financially supported by  RFFI Grant No.\ 96-02-17757.
E.V.S.\ wishes to thank prof. K.~Ishikawa for the kind hospitality 
during his visit to the Department of Physics, Hokkaido University,
Sapporo, Japan, where  part of the work was done.
A.V.Kh. thanks Dr. Yu.~Nazarov for useful discussion of the results of
this work.

\begin{figure}
\caption{The tunnel geometry. The magnetic field vector constitutes angle $ \theta$ with the interface plane.}
\label{1}
\end{figure}

\begin{figure}
\caption{Diagrams which contribute in the lowest order to two-particle Green's function 
Eq.(4)
are shown. Diagrams a) and c) describe correspondingly exchange and Hartree
corrections to one-particle Green's function in each lead. Diagram b) describes
the interaction of electrons in different leads. Wavy line is dynamically
screened Coulomb interaction. Diffusion vertexes are shown by shaded regions.}
\label{2}
\end{figure}

\end{document}